\documentclass[12pt]{article} 
\usepackage{graphicx,a4}
\newcommand{\preprintno}[1]
{\vspace{-2cm}{\normalsize\begin{flushright}#1\end{flushright}}\vspace{1cm}}

\title{\preprintno{{\bf SUSX-TH-99-013}}
Destabilizing the gaugino condensate in modular invariant supergravity}
\author{Thomas Dent\thanks{E-mail address: t.e.dent@sussex.ac.uk} \\ 
	{\em Centre for Theoretical Physics, University of Sussex,} \\
	{\em Brighton BN1 9QH, U.K.}}
\date{September 1999}

\begin{document}
\maketitle
\begin{abstract}
\noindent We investigate the stability of the hidden sector gaugino condensate in a SL(2,{\bf Z})-invariant 
supergravity model inspired by the E$_8\otimes$E$_8$ heterotic string, using the chiral 
superfield formalism. We calculate the Planck-suppressed corrections to the ``truncated approximation'' 
for the condensate value and the scalar potential. A transition to a phase with zero condensate occurs 
near special points in moduli space and at large compactification radius. We discuss the implications 
for the T-modulus dependence of supersymmetry-breaking.
\end{abstract}
\section{Introduction}
Dynamical supersymmetry-breaking via gaugino condensation \cite{Nilles+FerraraGN} in a hidden sector is a 
major component of realistic supergravity unified theories. In heterotic string theory with gauge 
group E$_8\otimes$E$_8$, the second E$_8$ factor provides a suitable hidden sector 
\cite{DerenIN+DineRSW+Taylor85} with one or more confining gauge groups (depending on the details of 
symmetry-breaking). Assuming weak coupling at unification, the gauge coupling will become strong at an 
energy scale\footnote{We use reduced Planck units with $\kappa^{-1} =1/\sqrt{8\pi G} =1.$} $\Lambda \sim 
e^{-1/2bg_X^2}$, where $g_X$ is the unified gauge coupling and $b$ is the one-loop beta function coefficient 
defined such that $\beta(g) = -bg^3+\ldots$ . The vacuum expectation value of the gaugino bilinear $\langle 
\mbox{Tr} \lambda^{\alpha} \lambda_{\alpha} \rangle$ switches on with a value of the order of $\Lambda^3$, 
breaking local supersymmetry. Supersymmetry-breaking is mediated by the dilaton and moduli superfields $S$, 
$T_i\,$ in the four-dimensional supergravity effective theory: the auxiliary fields $F^S$ and $F^{T_i}$ take 
values of the order of $\langle \lambda \lambda \rangle$, the flat scalar potential for $S$ and $T_i$ is 
lifted and supersymmetry is broken softly in the visible sector \cite{deCarlosCM93,BrignoleIM+KaplunovskyL93}. 

It has has usually been assumed that the condensate is well described by the {\em globally\/} 
supersymmetric Yang-Mills theory, since the confinement scale is well below $M_P$. In the limit of global 
supersymmetry $M_P \rightarrow \infty$, the condensate does not break supersymmetry 
\cite{VenezianoY,WittenIndex}: its value is determined in the effective superpotential approach 
\cite{VenezianoY}, where the gaugino bilinear is the lowest component of the composite chiral superfield 
$U= \mbox{Tr} W^\alpha W_\alpha$, by setting the auxiliary field $F_U= -(\partial \bar{W}_{np}/ \partial 
\bar{U})$ to zero, where $W_{np}$ is the nonperturbative superpotential generated by the gauge dynamics. 
The connection with supergravity models is via the value of the gauge coupling at unification, which is 
determined by the vacuum expectation values of the dilaton and moduli. The same condensate value follows 
from minimising the scalar potential in supergravity if higher-order terms in $U$ are neglected 
\cite{CasasLMR,LustT91}: we call this the ``truncated approximation''. The resulting condensate value is 
substituted back into the nonperturbative superpotential, which is then a function of the dilaton and 
moduli only and serves as an effective source of supersymmetry-breaking \cite{CasasLMR,deCarlosCM91,CveticFILQ}. 

If the gravitational corrections to the truncated approximation are significant, there will be effects on 
supersymmetry-breaking, on the stabilization of the dilaton and moduli, and on cosmological inflation models 
which use gaugino condensation to provide a positive potential \cite{BailinKL_inf,Thomas95+BanksBMSS}. 
The size of corrections is found by minimising the effective action, including gravitationally-suppressed 
terms, along the condensate direction. Lalak et al. \cite{LalakNN95} used a similar approach to take into 
account the ``backreaction'' of other fields on the condensate. Our analysis differs from theirs, since we 
assume a different mechanism for stabilizing the dilaton, we set all superpotential terms which do not depend on 
the condensate to zero, and we will be interested mainly in the effects on the compactification moduli of string 
theory, rather than on the dilaton.

\section{Effective action for the gaugino condensate} 
The vacuum expectation value (vev) of the gaugino bilinear is determined by the scalar component of $U$ 
at the stationary point of the effective action $\Gamma (U,S,T_i)$ as defined by Burgess et al. 
\cite{BurgessDQQ}, where $U$ is now the {\em classical\/} field which represents the expectation value 
of the composite operator Tr$(WW)$. For a single confining gauge group with no matter, the 
condensate is formally described by the field $U\equiv \hat{U}/S_0^3$, where $\hat{U} = 
\langle \mbox{Tr} W^{\alpha} W_{\alpha} \rangle$ and $W^\alpha$ is the gauge field strength chiral 
superfield. The chiral compensator superfield $S_0$ is introduced to simplify the formulation of 
supergravity coupled to matter \cite{SUGRA_matter} in the superconformal tensor calculus\footnote{The 
components of $S_0$ are determined by gauge-fixing the superconformal symmetries so that the Einstein 
term in the Lagrangian is canonically normalised \cite{BurgessDQQ}.}. The gaugino bilinear $\langle 
\mbox{Tr} \lambda^\alpha \lambda_\alpha \rangle$ is then the lowest component of $\hat{U}$ ($\theta= 
\bar{\theta}=0)$. 

The effective action $\Gamma (U,S,T_i)$ is constructed as a $\mbox{N}=1$ supergravity action, with a 
superpotential and K\"ahler potential which can be found by considering the symmetries of the underlying 
theory and the corresponding anomalies \cite{VenezianoY,BurgessDQQ}. In the case of a single overall modulus, 
$T_1= T_2= T_3\equiv T$, the superpotential is
\begin{equation}
	W_{np}(U,S,T)= \frac{1}{4}f_{GK}(S,T) U + \frac{b}{6}U \ln(cU) \label{eq:Wnp}
\end{equation}
where $f_{GK}(S,T)$ is the gauge kinetic function, equal to $S$ at tree level, which depends on the 
modulus $T$ through string loop threshold corrections \cite{DixonKL91}, and $c$ is a constant which will 
be discussed shortly. We can rewrite this as
\begin{eqnarray}
	W_{np}(U,S,T) &= &\frac{b}{6}U \ln(cU \omega(S)h(T)) \label{eq:Wnp_h} \\
	&= & \frac{b}{6}U \left( \ln \left(\frac{U}{\Lambda_c^3} \right) -1 \right) 
\label{eq:lambdac}
\end{eqnarray}
where $\omega(S)\equiv e^{3S/2b}$ and $h(T)$ is a function with appropriate transformation properties 
under the SL(2,{\bf Z}) target-space duality group \cite{FerraraMTV,Taylor90}. The condensation scale 
$\Lambda_c$ is defined to be
	\[ \Lambda_c \equiv e^{-S/2b-1/3}(c h(T))^{-1/3}. \]
Changing the value of $c$ is equivalent to changing the unified gauge coupling or the beta-function. In 
principle $c$ could be determined, if $f_{GK}$ is known, by comparing the resulting condensation scale 
with the vev of $\mbox{Tr} (W^\alpha W_\alpha)$ derived from instanton calculations in global SUSY 
\cite{Amati88}(compare \cite{FerraraMTV}).

The K\"ahler potential was determined in \cite{BurgessDQQ} as 
\begin{equation}
	K(U,S,T) = \tilde{K} - 3\ln\left(1- \frac{9}{\gamma}e^{\tilde{K}/3} (U\bar{U})^{1/3}\right) 
\label{eq:K1} 
\end{equation} 
where
\begin{equation}
	\tilde{K} = K_p(S,T) = -\ln(S+\bar{S}) -3 \ln(T+\bar{T}) \label{eq:Ktilde0}
\end{equation} 
is the perturbative string tree-level K\"ahler potential, and $\gamma$ is an real constant of order 1; we 
will show that the value of $\gamma$ does not affect our results. This expression for $\tilde{K}$ is also 
valid at one loop if the Green-Schwarz anomaly cancellation coefficients $\delta_{GS}$ are neglected: we 
set $\delta_{GS}=0$ for simplicity. The expression (\ref{eq:Ktilde0}) can be generalised to include 
stringy nonperturbative effects \cite{Shenker90} which have been invoked in order to stabilise the 
dilaton \cite{BanksDine+Casas96+BGW+ChoiKK}: we replace (\ref{eq:Ktilde0}) by 
	\[ \tilde{K} = P(y)- 3\ln(T+\bar{T}). \]
where $P(y)$ is a real function of $y\equiv (S+\bar{S})$. The dilaton dynamics will enter in the effective 
action only through the auxiliary field $F^S$, so we do not need to specify the form of $P(y)$; however, we 
require $P''(y)>0$ for the dilaton kinetic term to have the right sign.

Note that $K$ becomes ill-defined for $(9/\gamma)e^{\tilde{K}/3}(U\bar{U})^{1/3} \rightarrow 1$. It is 
not surprising that the effective action for $U$ breaks down in this limit, since it means that the 
condensate forms at the string scale and the gauge group is strongly-coupled {\em at unification\/}. This 
limitation has implications for the behaviour of the condensate at large Re($T$), which will be discussed 
later.

The supergravity action is invariant under target-space SL(2,{\bf Z}) transformations 
\cite{FerraraLST,FerraraLT_2} if the superpotential transforms as a modular form\footnote{For a 
discussion of modular forms see \cite{Rankin} or the appendix of \cite{CveticFILQ}.} of weight $-3$. The 
modulus $T$ transforms as 
\begin{equation}
	T \rightarrow \frac{aT-ib}{icT+d} \nonumber
\end{equation}
where $a,b,c,d$ are integers satisfying $ad-bc = 1$. Then $U$ must transform as
\begin{equation} 
	U \rightarrow \zeta_U(icT+d)^{-3} U, \label{eq:Utransf} 
\end{equation} 
where $\zeta_U$ is a (field-independent) complex phase, and the function $h(T)$ transforms as $h(T) 
\rightarrow \zeta_U^{-1} (icT+d)^3 h(T)$. The $U$-dependent part of the K\"ahler potential is then 
modular invariant. In order to avoid singularities inside the fundamental domain of SL(2,{\bf Z}), $h(T)$ 
must be of the form 
\begin{equation}
	h(T)= \eta^6(T)/H \label{eq:hH}
\end{equation}
where $\eta(T)$ is the Dedekind eta function and $H$ is a constant or a modular 
invariant function of $T$ without singularities \cite{CveticFILQ}.

The part of $\Gamma(U,S,T)$ relevant to finding the value of the condensate is the scalar potential, 
which is given as usual by
	\[ V = e^K((K_i \bar{W} + \bar{W}_i) (K^{-1})^i_j (K^j W + W^j) - 3|W|^2). \]
The indices $i$ and $j$ range over the scalar components of $U$, $S$ and $T$ (from now on these symbols will
denote the scalar components rather than the superfields), $X^i\equiv \partial X/\partial \phi_i$ and 
$X_i\equiv \partial X/\partial \bar{\phi}^i$ (where X is any function of the scalars and their 
complex conjugates), and $K^{-1}$ is defined by $(K^{-1})^i_j K^j_k = \delta^i_k$. We find, as in 
\cite{Taylor90,deCarlosCM91}:
\begin{equation} 
	V = \left(\frac{b}{6}\right)^2 \frac{|z|^4}{(9/\gamma)^2 (1-|z|^2)^2} \left\{3\left| 
1+\ln(c\omega(S)h(T) e^{-\tilde{K}/2} z^3)\right|^2 + C_2 |z|^2 \right\} \label{eq:VSTZ}
\end{equation}
where 
	\[ z= \frac{3}{\sqrt{\gamma}}(e^{\tilde{K}/2}U)^{1/3} = \frac{3}{\sqrt{\gamma}}\, e^{P(y)/6} 
(T+\bar{T})^{-1/2} U^{1/3}, \]
\begin{equation}
	C_2 = \frac{1}{P''(y)} \left|P'(y)- \frac{\omega'(S)}{\omega(S)} \right|^2 +\frac{1}{3} \left|3+ 
(T+\bar{T}) \frac{h'(T)}{h(T)} \right|^2 -3, \label{eq:C2}
\end{equation}
and we have absorbed a factor of $(9/\gamma)^{-3/2}$ into the constant $c$. The only dependence on $\gamma$ 
of the potential (\ref{eq:VSTZ}) is in the overall scale, which has no effect on the existence or stability 
of minima, so we set $9/\gamma=1$ from now on.

The potential depends on the phase of $z$ only through the first term inside the brackets. If we 
hold $|z|$ fixed and vary the phase, a minimum can only occur when the argument of the logarithm 
$(c\omega(S)h(T) e^{-\tilde{K}/3} z^3)$ is real and positive. The condensate phase $\arg(z)$ is then 
aligned with $\arg(h(T)^{-1/3})$ \cite{FerraraMTV}, for $c$ and $\omega(S)$ real. The 
dependence of the potential on $|z|$ is then determined by two real parameters, $|c\omega(S)h(T)| 
e^{-\tilde{K}/2}$ and $C_2$.

The truncated approximation amounts to minimising (\ref{eq:VSTZ}) in the rigid supersymmetry limit where 
$M_P \rightarrow \infty$, $\omega(S) \rightarrow \infty$, $z \rightarrow 0$ and $C_2$ is held constant. The 
first term inside the bracket then dominates the dependence on $z$ and there is a zero-value minimum 
with unbroken supersymmetry satisfying $\partial W/\partial U=0$ at 
\begin{equation} 
	z_{tr} = e^{\tilde{K}/6}\Lambda_c =\frac{e^{P(y)/6 -1/3}}{(T+\bar{T})^{1/2}} 
(c\omega(S)h(T))^{-1/3}. \label{eq:Ztr}
\end{equation}
Substituting this value back into (\ref{eq:Wnp_h}) leads to the well-known form of nonperturbative 
superpotential \cite{FontILQ90,CveticFILQ,deCarlosCM91} 
\begin{equation}	
	W_{np}(S,T) = \frac{b}{6} \Lambda_c^3 \propto \frac{\omega^{-1}(S) H}{\eta^6(T)} \label{eq:W_HT}
\end{equation}
which has been argued to occur independently of any particular super\-sym\-metry-breaking mechanism 
\cite{FerraraLST,CveticFILQ}.

\section{Beyond the truncated approximation}
Returning to local supersymmetry, when $C_2|z|^2$ is non-zero, $z=z_{tr}$ is not a minimum of the 
scalar potential and we cannot invoke the condition that the hidden sector gauge dynamics preserve 
supersymmetry. For small $C_2|z|^2$, the truncated value is a good approximation; however, we would like 
to find the size of corrections, and to investigate the behaviour of the potential when the truncated 
approximation fails.

Figure 1 shows the effect on the potential as a function on $|z|$ of varying $C_2$ while keeping 
$|z_{tr}|$ fixed. For \mbox{$-3<C_2<0$} there is an absolute minimum with $|z|$ near the truncated value 
and negative vacuum energy. For $C_2$ small and positive there is a local minimum at $|z|$ near 
$|z_{tr}|$ and the vacuum energy is positive; as $C_2$ increases the minimum becomes shallower and moves 
towards smaller values of $|z|$; finally the minimum merges with a point of inflection and the condensate 
value goes discontinuously to zero, as in a first-order phase transition\footnote{Ferrara et al. \cite{FerraraMTV} 
already noticed that for some values of parameters the only stationary point of the 
potential is at $z=0$.}.

\begin{figure}[tb]
\centering
\includegraphics[width=10.5cm,height=8.5cm]{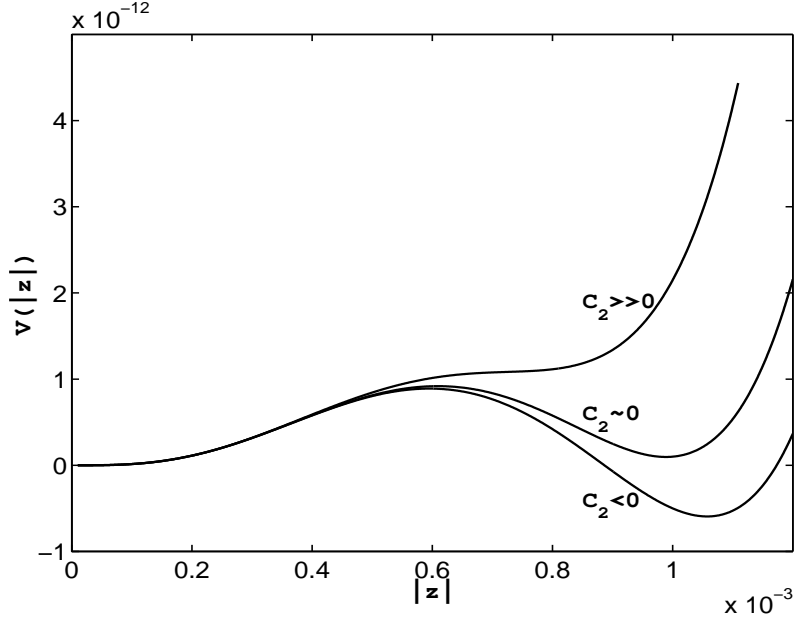}
\caption{Effect of varying the parameter $C_2$ on the potential $V(|z|)$, setting $|z_{tr}|=10^{-3}$.}
\label{fig:varyC2}
\end{figure}

The stationary point condition for the potential (\ref{eq:VSTZ}) is a transcendental equation in $z$, so 
we use a series expansion for $z$ near the truncated value and also search for stationary points 
numerically. We rewrite the potential as
\begin{equation} 
	V(|z|) = \left(\frac{b}{6}\right)^2 \frac{|z|^4}{(1-|z|^2)^2} \left\{ 3\left(\log 
\left|\frac{z}{z_{tr}} \right|^3 \right)^2 +C_2|z|^2 \right\}. \label{eq:VZtrC2} 
\end{equation}
and expand to third order in $\Delta z \equiv |z|-|z_{tr}|$, obtaining
\begin{equation} 
	V = \left(\frac{b}{6}\right)^2 \frac{|z_{tr}|^4}{(1-|z_{tr}|^2)^2} \left( C_2|z_{tr}|^2 
+a_1\Delta z +a_2\Delta z^2 +a_3\Delta z^3 + \ldots \right) 
\label{eq:Vorder3} 
\end{equation}
where
\begin{eqnarray*} 
	a_1 & = & 2C_2|z_{tr}|(1-|z_{tr}|^2)(3-|z_{tr}|^2) \\
	a_2 & = & 27\frac{(1-|z_{tr}|^2)}{|z_{tr}|^2} +C_2(15-4|z_{tr}|^2+|z_{tr}|^4) \\
	a_3 & = & 27\frac{(1-|z_{tr}|^2)(3+|z_{tr}|^2)}{|z_{tr}|^3} + 4C_2 \frac{(5 
+3|z_{tr}|^2)}{|z_{tr}| (1-|z_{tr}|^2)}.
\end{eqnarray*}
Since $|z_{tr}|$ is ${\cal O}(10^{-1})$ or smaller in most cases of physical interest, we neglect 
$|z_{tr}|^2$ next to 1 and take
\begin{equation} 
	a_1= 6C_2|z_{tr}| \mbox{,  } a_2= \frac{27}{|z_{tr}|^2} +15C_2 \mbox{,  } a_3= 
\frac{81}{|z_{tr}|^3} +\frac{20C_2}{|z_{tr}|}. \label{eq:a123}
\end{equation}

The series expansion (\ref{eq:Vorder3}) has stationary points if $a_2^2 \geq 3a_1a_3$, with a stationary 
point of inflection if the equality is satisfied. In terms of the parameter $x \equiv C_2|z_{tr}|^2$, 
the condition for a minimum to exist is
\begin{equation} 
	27 -24x -5x^2 >0 \label{eq:mincond}
\end{equation} 
which is solved, for positive $x$, by $x<0.94$. 

Where (\ref{eq:mincond}) holds, the minimum is at
\begin{equation}
	\Delta z_{3} = \frac{1}{3a_3} \left(-a_2 + \sqrt{a_2^2 - 3a_1a_3} \right) = |z_{tr}| 
\frac{\left(-9-5x +\sqrt{81 -72x -15x^2} \right)}{81+20x} \label{eq:Zm123}
\end{equation}
which can be expanded as $\Delta z_{3}= |z_{tr}|(-x/9+x^2/162 +\ldots)$ for small x. The leading order 
corrections to the potential at its minimum in the $z$-direction are 
	\[ V(|z|_3) - V(z_{tr}) = \frac{b^2}{36} \frac{|z_{tr}|^4}{(1-|z_{tr}|^2)^2} \left( 
-\frac{x^2}{3} +\frac{2x^3}{27} +\ldots \right) = V(z_{tr}) \left(-\frac{x}{3} + \frac{2x^2}{27} +\ldots \right) \]
where $|z|_3=|z_{tr}| +\Delta z_{3}$.

\begin{figure}[tb] 
\centering
\includegraphics[height=8.5cm,width=10.5cm]{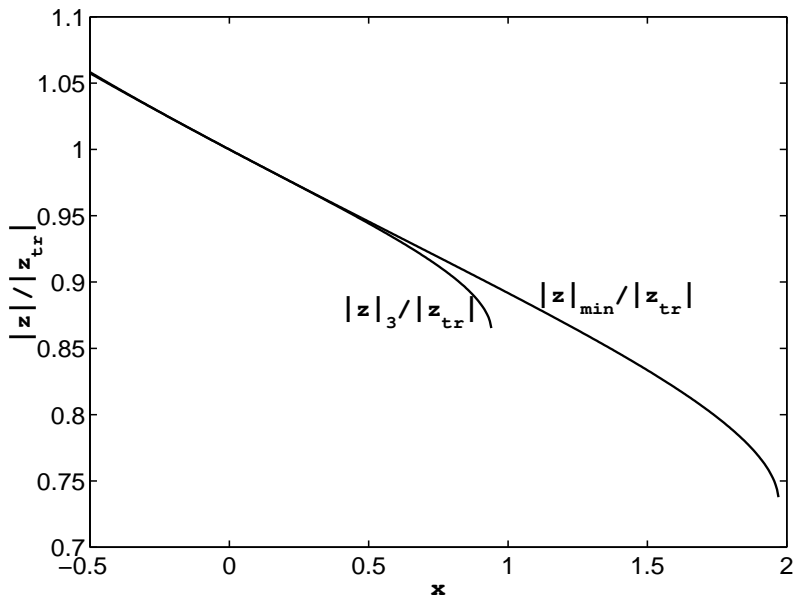}
\caption{Values of $|z|$ at the minimum as a function of $x \equiv C_2|z_{tr}|^2$, scaled by 
$|z_{tr}|=10^{-3}$. $|z|_{min}$ and $|z|_3$ minimise the full potential (\ref{eq:VZtrC2}) and the 
third-order series expansion, respectively.}
\label{fig:ZmZm3}
\end{figure}

We also found the condensate value $|z|_{min}$ at the minimum of the full potential (\ref{eq:VZtrC2}) 
numerically, as a function of $x$. For all values of $|z_{tr}|<2\times 10^{-2}$, the phase transition at 
which $|z|_{min}$ goes to zero occurs for $x$ between 1.968 and 1.972. For larger $|z_{tr}|$, the 
transition occurs at slightly smaller values of $x$, indicating a dependence on higher-order terms in 
$|z_{tr}|$: for example, at $z_{tr} =0.2$ the transition occurs for $1.872 <x< 1.876$. Figure 
\ref{fig:ZmZm3} shows the value of $|z|/|z_{tr}|$ at the minimum for the full potential and for the 
series expansion (\ref{eq:Vorder3}), as a function of $x$.
We take $|z_{tr}| = 10^{-3}$; however, the results are very similar for all values of $|z_{tr}|$ below 
about $0.3$. Both curves differ significantly from 1, and the series expansion is a good approximation 
to the true minimum for $x<0.7$. While the series expansion gives a qualitative picture of the behaviour 
of $|z|$ near the phase transition, the value of $x$ where the transition occurs is about twice that 
predicted by (\ref{eq:mincond}). This is to be expected, since $\Delta z$ is large here and the series 
expansion is less accurate. 

\begin{figure}[tb] 
\centering{\includegraphics[height=9cm,width=10cm]{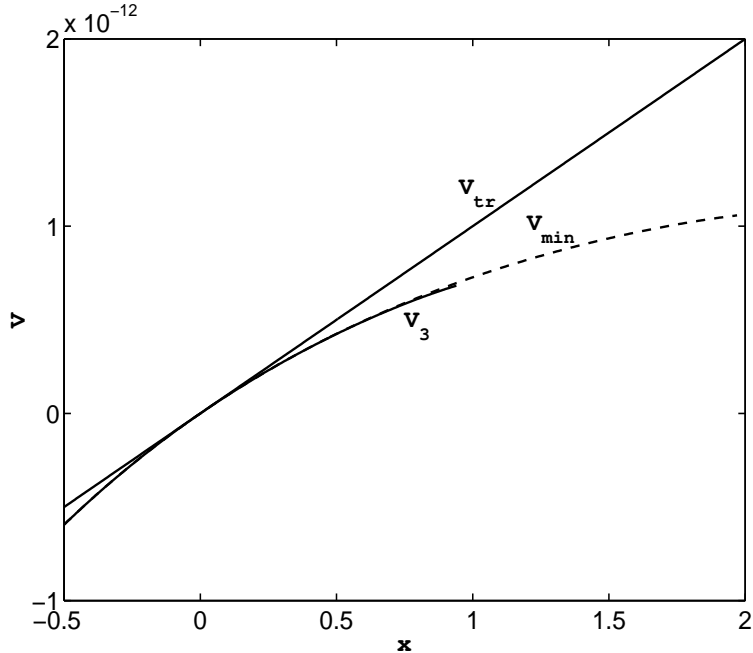}}
\caption{$V_{tr}$ and $V_{min}$ are the value of the potential at $|z_{tr}|$ and at $|z|_{min}$, respectively. 
$V_3$ is the value of the third-order series expansion (\ref{eq:Vorder3}) at its minimum 
$|z|_3$ ($|z_{tr}| =10^{-3}$).}
\label{fig:V_atZm}
\end{figure}

Figure \ref{fig:V_atZm} shows the value of the potential at its minimum and at the truncated value of $z$, 
and the minimum value of the series expansion, as a function of $x$. 
Again, there is a significant deviation from the truncated result for $x \geq$ few$\times 0.1$.
 
\section{Discussion}
In what physical situations are the corrections to the truncated approximation significant? In order to 
generate phenomenologially reasonable supersymmetry breaking in the visible sector, we require the gravitino 
mass $m_{3/2}$ to be of the order of a TeV (see for example \cite{deCarlosCM93}); for a single condensate it is 
given by $m_{3/2} \simeq \frac{b}{6} |z|^3$, which implies a vev of $|z|_0 = \mbox{few}\times 10^{-5}$ for the 
physical values taken by the dilaton and moduli today. The parameter $x= C_2|z_{tr}|^2$ should be at least 
${\cal O}(0.1)$ for corrections to the truncated approximation to be appreciable, so unless one of the terms 
in $C_2$ is of order $10^{9}$ or larger the truncated approximation is not in any danger today. 

The first term in (\ref{eq:C2}) can become large only if the ratio $P'(y)^2/P''(y)$ does (note that 
$\omega'/\omega$ is a constant of order 10). It is unlikely that the function $P(y)$, which describes 
stringy non-perturbative effects on the dilaton, will give rise to singularities or very large numbers; 
in addition, since $P''(y)$ cannot change sign, it cannot be very small without fine-tuning. The only limit in 
which the dilaton-dependence causes the truncated approximation to break down is at strong string coupling ($S 
\rightarrow 0$), where the effective action is anyway not valid. 

The second term involving $h'(T)/h(T)$, where $h(T)$ is given by (\ref{eq:hH}), may become large if 
$H(T)$ has zeros or singularities in the fundamental domain; in addition, the eta function 
decays exponentially with $T$ at Re$(T)\rightarrow \infty$, which leads to a divergence at large 
Re$(T)$\footnote{This behaviour may be understood in terms of string states charged under the gauge group 
which become light (mass $\ll M_P$) at certain values of $T$ \cite{CveticFILQ}. At large Re$(T)$ the 
extra dimensions decompactify and an infinite number of Kaluza-Klein states become light, producing a 
linear divergence in the gauge kinetic function. At poles or zeros of $H$ a finite number of states 
become light, leading to a logarithmic divergence in $f_{GK}$.}. Nothing can be said within the effective 
theory about the limit of large Re$(T)$ or the singularities of $H(T)$, since the condensation scale 
$\Lambda_c$ diverges at these points. However, at zeros of $H(T)$, $f_{GK}$ becomes large and positive and 
the condensation scale goes to zero, so the effective action can be used to describe the condensate at 
these points. 

Consider the ansatz for the $T$-dependence of the superpotential (\ref{eq:Wnp_h},\ref{eq:hH}) 
involving the absolute modular invariant $J(T)$ \cite{CveticFILQ}, where $H(T)$ takes the form
\begin{equation}
	H(T) = (J-1)^{m/2}J^{n/3}p(J) \label{eq:H_J}
\end{equation}
where $m$, $n$ are positive integers or zero and $p(J)$ is a polynomial in $J$. This form for $H$ has no 
singularity in the fundamental domain, but may have zeros at the ``fixed point'' values $T= \rho \equiv 
e^{i\pi/6}$ and $T=1$, since $J\propto (T-\rho)^3$ near $T=\rho$ and $J\propto (T-1)^2$ near $T=1$. 

As $T$ approaches a zero of $H(T)$ (we consider $T=\rho$ for definiteness), $|z_{tr}|$ vanishes with 
$|T-\rho|^{n/3}$ and $C_2$ diverges with $|T-\rho|^{-2}$. The potential in the truncated approximation 
$V_{tr} = (b/6)^2 |z_{tr}|^6 C_2$ varies smoothly as $T \rightarrow \rho$; in the case $n=1$, the 
potential tends to a non-zero value, even though the condensate vanishes at $T=\rho$ and supersymmetry is 
unbroken. This puzzle is resolved by looking at the stabilization of the condensate explicitly: as $T 
\rightarrow \rho$, the parameter $x$ diverges as $x \propto |T-\rho|^{2n/3-2}$, so for $1\leq n<3$ the 
truncated approximation breaks down and $|z|_{min}$ goes to zero abruptly at some finite, small value of 
$|T-\rho|$. The scalar potential as a function of $T$ develops a ``hole'' around $T=\rho$, inside which 
the condensate and scalar potential vanish and supersymmetry is restored. If $T$ were to fall into such a 
region, since $z=0$ is always a stationary point of the effective action, a non-zero condensate would not 
immediately be restored if $T$ then rolled out to a value where (\ref{eq:mincond}) held: the universe 
would have to ``tunnel'' back to a supersymmetry-breaking vacuum.

We also consider varying $T$ away from its present-day value in a perturbative heterotic string vacuum 
with $S\simeq 2$ and $T_0 = {\cal O}(1)$, to find the effect of our treatment of the condensate on the 
(cosmological) evolution of the moduli if Re$(T)$ becomes large. As was argued above, the present-day 
value of the condensate should be $|z|_0 \simeq 2\times 10^{-5}$ in Planck units. In the case where 
$h(T) = \eta^6(T)$, $z_{tr}$ varies as $\eta^{-2}(T)$: for Re$(T)>3$ this is well approximated by 
$\eta^{-2}(T) = e^{\pi T/6}$, while the relevant part of $C_2$ is given by $C_2 \simeq 
\pi^2(T+\bar{T})/12 \simeq 1.64$ Re$(T)$. The effective theory breaks down when $|z_{tr}|$ approaches 
$1$, which occurs when $T \simeq 21$; however, a non-zero condensate is only stable for $1.64\, T e^{\pi 
T/3} (2\times 10^{-5})^2 < 1$ (see eqn. \ref{eq:mincond}), which implies that $T< 17$--$19$ (depending on 
the exact values of $|z_{tr}|_0$ and $\eta(T_0)$). At the phase transition where the condensate collapses, 
$C_2 \simeq 30$ so $|z_{tr}| \simeq 0.18$, which is consistent with using the effective action.

This analysis can also be applied to models of cosmological inflation which use the $T$ modulus to 
provide a positive vacuum energy \cite{BailinKL_inf,Thomas95+BanksBMSS}. In the region of the complex $T$ 
plane where inflation occurs, $V\simeq |z_{tr}|^6 C_2$ should be of order $10^{-10} \kappa^{-4}$ to 
produce the correct amplitude of CMB fluctuations at large angular scales \cite{Thomas95+BanksBMSS}. 
Typically, this can be achieved for $T= {\cal O}(1)$ and $C_2 = {\cal O}(10)$, implying that $|z_{tr}|$ 
is of order $10^{-2}$ and $x\sim 10^{-3}$: the condensate scale here is much larger than that required to 
give realistic SUSY-breaking. Repeating the previous analysis, we find that the effective theory becomes 
invalid at Re$(T) \simeq 9$ and a non-zero condensate ceases to be stable for Re$(T)$ between $6.5$ and $7$.

One might hope to find a form of $T$-dependent superpotential (\ref{eq:hH}) involving $J(T)$ 
(\ref{eq:H_J}) which allows $T$ to take large values while keeping modular invariance of the effective 
action, since it has been suggested that a realistic vacuum of the strongly-coupled heterotic string 
\cite{HoravaWitten} may have Re$(T)$ of order $25$ \cite{BanksDine96+}. Unfortunately the first term in 
the large $T$ expansion of $J(T)$ increases as $e^{2\pi T}$, so the exponential growth of $z_{tr} \propto 
H^{1/3}\eta^{-2}$ at large Re$(T)$ is likely to be faster than for constant $H$, and there is little prospect 
of stabilizing either the condensate or $T$ with this ansatz.

\section{Conclusions}
We have studied the stability of a non-zero gaugino condensate in an effective supergravity model 
motivated from compactifications of heterotic string theory with duality group SL(2,{\bf Z}). We defined a 
parameter $x$ which determines the deviation of the value of the condensate from its value in global 
supersymmetry. Using a series expansion for the effective potential, the corrections to the truncated 
approximation at small $x$ were obtained to second order in $(\Lambda_c/M_P)^2$ for the value of the condensate, 
and to first order in $(\Lambda_c/M_P)^2$ for the vacuum energy as a function of the dilaton and $T$ modulus. 
Numerical minimisation of the effective potential confirms the series expansion results at small $x$, and 
reveals a phase transition at large $x$ where the condensate vanishes discontinuously. This appears to be 
caused by the backreaction of the dilaton and $T$ modulus on the hidden sector gauge dynamics. There are 
implications for the $T$-dependence of hidden sector supersymmetry-breaking : the auxiliary field $F^T$ and the 
scalar potential receive large corrections or go to zero abruptly near special points in moduli space and at 
large Re$(T)$. In particular, $T$ is not stabilized against arbitrarily large fluctuations.

It would be interesting to find the effect of non-zero Green-Schwarz anomaly cancellation coefficients on 
the $S$- and $T$- dependence of the scalar potential, and also to analyse the stability of the condensate in 
the linear supermultiplet formalism \cite{BinGWu}. If the potential for $T$ is very flat near the minimum, 
it may be worth investigating the effect of including Planck-suppressed corrections on the $T$-dependent 
Yukawa couplings and soft supersymmetry-breaking terms, particularly if CP-violating phases depend strongly 
on the vev of $T$ \cite{BailinKL_CP}.

\section*{Acknowledgements}
Thanks are due to David Bailin for suggesting the problem, to Jackie Grant for help with the figures and to 
Malcolm Fairbairn for a critical reading of the paper in its early stages. TD is supported by PPARC 
studentship PPA/S/S/1997/02555.

\end{document}